\documentclass[12pt,a4paper]{article}

\usepackage{latexsym}
\usepackage{setspace}
\usepackage{a4wide}
\usepackage{amsmath}
\usepackage{amsfonts}
\usepackage{amscd}
\usepackage{amssymb}
\usepackage{amsbsy}
\usepackage{mathrsfs}
\usepackage{yfonts}
\usepackage{bbm}
\usepackage[dvips]{epsfig,graphics}
\usepackage{epsfig}
\usepackage{graphicx}
\usepackage{psfrag}
\usepackage{upgreek}
\usepackage{undertilde}
\usepackage{color}
\usepackage{leftidx}
\usepackage{mathbbol}

\DeclareMathAlphabet{\mathpzc}{OT1}{pzc}{m}{it}

\begin{document}

{\color{red}{This document is the Accepted Manuscript version of a Published Work that appeared in final form in Journal of Applied Physics, copyright American Institute of Physics after peer review and technical editing by the publisher. To access the final edited and published work see \\ http://aip.scitation.org/doi/full/10.1063/1.4928464}}

\title{Graphene as a hexagonal 2-lattice: evaluation of the in-plane material constants for the linear theory. A multiscale approach.}

\author{D. Sfyris, E.N. Koukaras, N. Pugno, C. Galiotis}

\maketitle

\begin{abstract}

Continuum modeling of free-standing graphene monolayer, viewed as a two dimensional 2-lattice, requires specification of the components of the shift vector that acts as an auxiliary variable. If only in-plane motions are considered the energy depends on an in-plane strain measure and the shift vector. The assumption of geometrical and material linearity leads to quadratic energy terms with respect to the shift vector, the strain tensor, and their combinations. Graphene's hexagonal symmetry reduces the number of independent moduli then to four. We evaluate these four material parameters using molecular calculations and the AIREBO potential and compare them with standard linear elastic constitutive modeling. The results of our calculations show that the predicted values are in reasonable agreement with those obtained solely from our molecular calculations as well as those from literature. To the best of our knowledge, this is the first attempt to measure mechanical properties when graphene is modeled as a hexagonal 2-lattice. This work targets at the continuum scale when the insight measurements comes from finer scales using atomistic simulations. 

\end{abstract}


\section{Introduction}

Ever since its discovery (\cite{Geim-Novoselov2007}) graphene attracted significant attention in the mechanics literature. Many works are devoted on evaluating graphene's Young's modulus and Poisson ratio, either by experimental or computational means. Lee et al (\cite{Leeetal2008}) conduct nanoidentation measurements using an atomic force microscope and measure Young modulus, E, of 340 $\pm$ 40 N m$^{-1}$, or of 1 $\pm$ 0.15 TPa when graphene's thickness is assumed to be 0.335 nm, for a monolayer graphene. Cadelano et al (\cite{Cadelanoetal2009}) combine tight binding atomistic simulations with continuum elasticity theory and report an E of 312 N m$^{-1}$ (0.93 TPa) and Poisson ratio $\nu=0.31$. Other tight binding claculations (\cite{Hernandezetal1998}) report an E of 1.21 TPa.

Zhou and Huang (\cite{Zhou-Huang2008}) utilize molecular dynamics and employ the Tersoff--Brenner potential to evaluate E=235 N m$^{-1}$ (0.70 TPa), and $\nu=0.413$. Zhou etal (\cite{Zhouetal2013}) use molecular mechanics to simulate an identation experiment end valuate E=1.19 TPa. A molecular dynamics method using the Brenner potential (\cite{Guptaetal2005}) render E=1.272 TPa and $\nu$=0.147, while Reddy et al (\cite{Reddyetal2006}) use the Tersoff--Brenner potential to arrive at E=0.67 TPa, $\nu=0.42$. 

Empirical force constant calculations (\cite{Michel-Verberck2008}) report E=384 N m$^{-1}$ (1.15 TPa), and $\nu=0.227$, while ab-initio calculations (\cite{Kudinetal2001}) arrive at E=345 N m$^{-1}$ (1.02 TPa), and $\nu=0.149$. Other works (\cite{Liuetal2007}) utilize ab-initio methods as well and report E=350 N m$^{-1}$ (1.04 TPA), and $\nu=0.186$. Kalosakas et al (\cite{Kalosakasetal2013}) perform calculations from first principles to parametrize classical potentials and evaluate a Young modulus of 320 N m$^{-1}$ (0.95 TPa). A density functional theory (\cite{Konstantinovaetal2006}) render E=1.24 TPa.  

Arroyo and Belytschko (\cite{Arroyo-Belytschko2004}) use a finite deformation continuum theory derived from interatomic potentials to derive E=235 N m$^{-1}$ (0.70 TPa), and $\nu=0.413$. Essentially, they use a finite element formulation whose potential derive from atomistic pictures, in line with the quasicontinuum approach (\cite{Tadmoretal1999}), combined with an appropriate definition of the Cauchy--Born rule for surfaces (\cite{Arroyo-Belytscko2002}). Finite element calculations using the truss model (\cite{Reddyetal2005}) result at E=1.11 TPa and $\nu=0.45$. The braced truss model (\cite{Scarpaetal2009}) using the AMBER force field result at E=1.22 TPa, while when the Morse force field is used they render E=1.91 TPa. In a recent review paper (\cite{Galiotisetal2015}) we summarize the relevant literature on graphene mechanics as probed by deformation and spectroscopic measurements and as calculated by ab-initio, molecular simulations and continuum mechanics methods.

The present work is a continuation of our previous efforts (\cite{Sfyris-Galiotis2015,Sfyrisetal2014a,Sfyrisetal2014b,Sfyrisetal2015}) to properly model graphene at the continuum level. The starting point is the modeling of graphene as a hexagonal 2-lattice (\cite{Fadda-Zanzotto2000}), in line with well established theories of multilattices (\cite{Ericksen1979,Parry1978,Pitteri1984,Pitteri1985,Pitteri-Zanzotto2003}). By making appropriate hypothesis (see \cite{Sfyris-Galiotis2015}) one works with an energy depending on an in-plane strain measure, the curvature tensor and the shift vector. Graphene's symmetry is taken into account, in this framework, by adding the structural tensor to the list of independent variables of the energy. This way the complete and irreducible representation of the energy is evaluated and from it, the stress tensor, the couple stress tensor as well as the driving force related with the shift vector. 

Simple closed form solutions for this genuinely geometrically and materially nonlinear theory are reported in \cite{Sfyrisetal2014a}. The geometrically and materially linear counterpart of the above theory is given in \cite{Sfyrisetal2014b}. There, graphene's energy is assumed to have quadratic dependence on the in-plane strain measure, the curvature tensor, the shift vector as well as to quadratic combinations of them. Hexagonal symmetry reduces then the overall number of moduli to nine. If in-plane motions are considered, only four material parameters should be determined; these are the constants $c_1, c_2, c_5, c_9$ in the terminology of \cite{Sfyrisetal2014b}. 

The present work is concerned with the evaluation of these four material parameters using molecular mechanics with the AIREBO potential. So, while the overall theory applies to the continuum scale, the calculations come from atomistic insights in finer scales. We correspond to the material parameters at the continuum level, four well defined measures from molecular considerations. The strategy for doing that is non-standard and goes as follows: we start at the discrete level where we focus on graphene's unit cell and distinguish between the measured length of the shift vector and the length of the lattice vectors. Then, we apply a tensile strain up to 6$\%$ along the armchair direction for a graphene monolayer that contains 31600 carbon atoms. Then, we evaluate the radial distribution diagram describing length change due to loading for carbon--carbon connections. 

At zero strain level, we find two peaks on the radial distribution diagrams: one corresponding to the equal length of the lattice vectors (approximately 0,242 nm), while the other peak corresponds to the shift vector (approximately 0,140 nm). As strain is gradually applied, we find that these peaks split into 2 new peaks each. These peaks measure changes that happen to graphene's unit cell due to applied strain. To these four peaks we correspond, at the continuum level, the four required material parameters $c_1, c_2, c_5, c_9$. To do this we first define four strain measures as the differences between the length at the peak point for strain level of 6$\%$, minus the initial length corresponding to the peak at zero strain, divided by the initial length. We plot the applied stress versus these four newly defined strain quantities. Slopes of these four diagrams correspond to the material parameters $c_1, c_2, c_5, c_9$.     

As a minimum validation/calibration of our approach we compare with reported values of E, and $\nu$ from the literature. To obtain the relation of $(E, \nu)$ with $(c_1, c_2, c_5, c_9)$ we solve the equations ruling the shift vector, to express the components of the shift vector as a function of the strain components. Since the problem is geometrically and materially linear this is feasible; for the nonlinear case this would have been cumbersome, if not non-solvable explicitly. Having these expressions at hand, we can invert the stress--strain relations to obtain the required expression of $(E, \nu)$ as function of $(c_1, c_2, c_5, c_9)$. Having $(c_1, c_2, c_5, c_9)$ evaluated from molecular calculations, we can then evaluate $(E, \nu)$=(1.37 TPa, 0.41) for our framework. 

Values for E and $\nu$ from the reported literature (see the first four paragraphs of this Section) range as E=0.67-1.91 TPa and $\nu$=0.14-0.45 depending on the methodology used. The central tendency of these values for E is the value 1 $\pm$ 0.15 TPa. Compared to this value our outcome of $(E, \nu)$=(1.37 TPa, 0.41) overestimates these quantities but still remain within the range of acceptable values. From the literature cited, the continuum methods (i.e. the finite element approaches of \cite{Arroyo-Belytschko2004,Reddyetal2005,Scarpaetal2009}) tend to have greater discrepancy from the value 1 $\pm$ 0.15 TPa. Thus, our theory, being ultimately a continuous theory, is expected to follow this trend. 

On the other hand, our pure molecular mechanics modeling render values $(E, \nu)$=(0.95 TPa, 0.20). These values are obtained using the definition of the AIREBo manual. But, the values $(E, \nu)$=(1.37 TPa, 0.41) are based on a different definition of $(E, \nu)$: they are based on a genuinely continuous definition which is non-standard since it uses $c_1, c_2, c_5, c_9$. Certainly, the two definitions (the discrete and the continous one) measure the same quantities in a different way. So, the discrepancy in their reported values is based on the different definition but still remains in the range of an admissible difference.  

The paper is organized as follows. Section 2 presents compactly the key findings of our previous works (\cite{Sfyris-Galiotis2015,Sfyrisetal2014a,Sfyrisetal2014b}), to which we refer for more information. In Section 3 we present the core of our caclulations. We lay down the strategy for obtaining/defining the required material parameters at the continuum level starting from discrete pictures and measurements using the AIREBO potential. Section 4 gives the minimum validation by correlating with standard results. The article ends up in Section 5 with some concluding remarks as well as future directions. As far as notation is concerned, we use tensor notation in component form throughout the paper. All indices are assumed to refer to the same Cartesian coordinate system and range from 1 to 2.  
   
\section{Modeling of graphene as a 2-lattice: the linear case}

We begin by presenting the main findings of our relevant previous works (\cite{Sfyris-Galiotis2015,Sfyrisetal2014a,Sfyrisetal2014b}) that constitute the theoretical backbone of this work. Some more detailed information regarding the continuum modeling of crystalline materials can be found in the excellent book by Pitteri and Zanzotto (\cite{Pitteri-Zanzotto2003}). 

Graphene is modeled as a hexagonal 2-lattice (\cite{Fadda-Zanzotto2000}) at the crystalline level. One arrives at the continuum level by assuming validity of the Cauchy--Born rule (\cite{Ericksen2008}). Validity of this rule, together with confinement to weak transformation neighborhoods (\cite{Pitteri1984,Pitteri1985}) enables one to work with the symmetry groups classical elasticity uses; for the case of graphene these are rotations by 60$^0$. 

For the geometrically and materially linear case energy depends on the in-plane strain tensor
\begin{equation}
e_{\alpha \beta}=\frac{1}{2}(u_{\alpha, \beta}+u_{\beta , \alpha}),
\end{equation}
$u$ being the in-plane displacement, the curvature tensor $\bf b$ as well as the shift vector $\bf p$. Explicitly, energy has the form (\cite{Sfyrisetal2014b})
\begin{eqnarray}
W({\bf e}, {\bf b}, {\bf p})&&=\frac{1}{2} C^1_{ijkl} e_{ij} e_{kl} +\frac{1}{2} C^2_{ij} p_i p_j +\frac{1}{2} C^3_{ijk} e_{ij} p_k \nonumber\\
&&+\frac{1}{2} C^4_{ijkl} b_{ij} b_{kl} +\frac{1}{2} C^5_{ijkl} e_{ij} b_{kl} +\frac{1}{2} C^6_{ijk} b_{ij} p_k. 
\end{eqnarray} 
Tensors ${\bf C}^1, ...,{\bf C}^6$ are tensors of material moduli. When out-of-plane motions are neglected, terms related with the curvature should be set equal to zero. In this case the stress--strain relations read (\cite{Sfyrisetal2014b})
\begin{eqnarray}
\sigma_{11}&&=c_1 e_{11}+c_2 e_{22} -c_5 p_2, \\
\sigma_{22}&&=c_2 e_{11} + c_1 e_{22}+c_5 p_2, \\
\sigma_{12}&&=\frac{c_1-c_2}{2} e_{12}-2 c_5 p_1,
\end{eqnarray}
steming from the expression for the stress tensor
\begin{equation}
{\boldsymbol \sigma}=\frac{\partial W}{ \partial {\bf e}}=C^1_{ijkl} e_{kl}+C^3_{ijk} p_k.
\end{equation}

For the components related with the shift vector we have
\begin{equation}
\frac{\partial W}{\partial p_i}=C^2_{ij} p_j+C^3_{ijk} e_{jk}.
\end{equation}
So, we finally take
\begin{eqnarray}
\frac{\partial W}{\partial p_1}=&&c_9 p_1-2 c_5 e_{12}, \\
\frac{\partial W}{\partial p_2}=&&c_9 p_2 -c_5 e_{11}+c_5 e_{22}.
\end{eqnarray}

The field equations for such a model are the momentum equation in the absence of body forces and inertial terms
\begin{equation}
\sigma_{ij,j}=0,
\end{equation}
and the equation ruling the shift vector
\begin{equation}
\frac{\partial W}{\partial p_i}=0.
\end{equation}
From the physical point of view the momentum equation is the force balance for the surface. The equation ruling the shift vector express that the shift vector adjusts so as equilibrium is reached. 

So, for the above framework we should evaluate material parameters $(c_1, c_2, c_5, c_9)$. These are material moduli at the continuum level which are present in the constitutive law, thereby characterizing graphene's mechanical properties in the small strain regime. Next section describes how these material parameters can be obtained/defined using molecular calculations. 

\section{Calculation of $c_1, c_2, c_5, c_9$ from molecular pictures}

For our purposes we load a graphene sheet along the armchair direction as Figure 1 shows. 
\begin{figure}[!htb]
\centering
\includegraphics{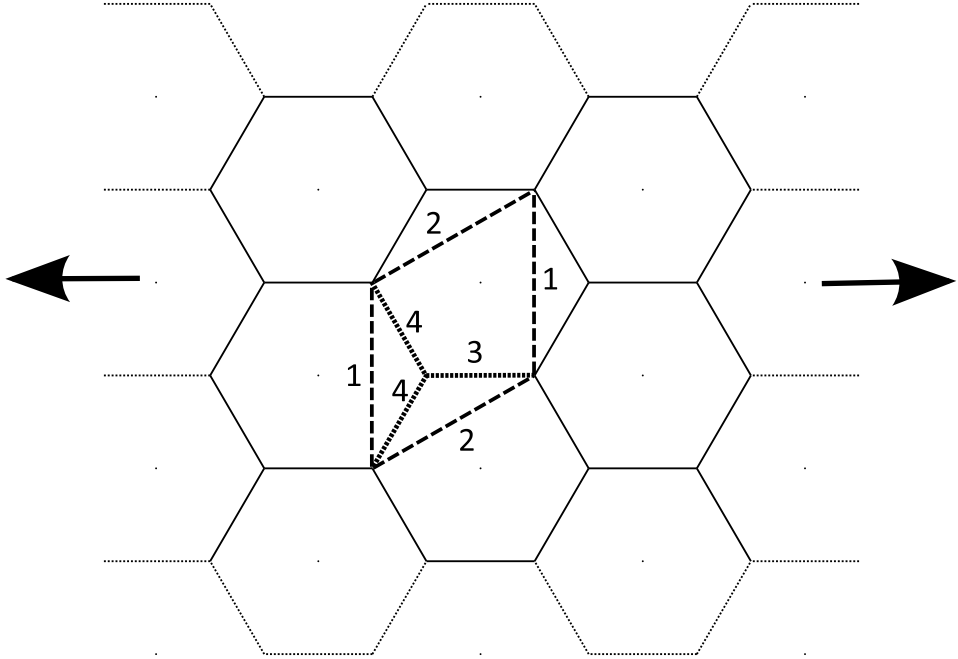}
\caption{ Unit cell and loading direction for graphene. Numbers 1 and 2 pertain to pairs of lattice vectors having equal length changes due to loading, while no. 4 pertains to a pair of the shift vector components having the same length change due to loading. No. 3 is another component of the shift vector which has different length change from no. 4.}
\label{fig:digraph}
\end{figure}
In this figure we depict the unit cell at ease, as well as the loading direction of the uniaxially strain which applies. Within the unit cell, we differentiate with numbers 1, 2, 3, 4 pair of carbon--carbon distances. At ease, pair of distances no. 1 have equal length of approximately 0.242 nm. The same holds true for the pair of distances denoted by no. 2. The pair of distances denoted by no. 4 have equal lengths of 0.140 nm, approximately. The same length is shared by distance denoted by no. 3 in Figure 1. Inspecting the unit cell of graphene when it is modeled as a 2-lattice (see figures in \cite{Sfyris-Galiotis2015,Sfyrisetal2014a,Sfyrisetal2014b}) it appears that lenghts no. 1, 2 correspond to the lattice vectors of graphene. Lengths numbered as no. 3, 4 correspond to graphene's shift vector. 

Now we apply an axial strain along the direction shown in Figure 1. This is done by employing the Adaptive Intermolecular Reactive Empirical Bond Order (AIREBO) (\cite{Stuartetal2000}) potential to describe the carbon--carbon interatomic forces. An orthogonal periodic computation cell is used with dimension 42.6 nm $\times$ 19.3 nm comprising of 31600 carbon atoms and the {\it x}-axis along the armchair direction (the direction of loading). The computational cell is initially relaxed, leading to an equilibrium structure for the given potential. The in-plane symmetry of the structure is broken by assigning randomized velocities with a Gaussian distribution to all of the atoms corresponding to a temperature of T = 40 K. The large number of atoms considered in the computational cell is needed to properly capture the distribution of the nearest and next nearest neighbor distances, in what follows. An energy equilibriation is performed within the microcanonical ensemble (NVE). The structure and unit cell are further relaxed by a follow up equilibriation within the isothermal--isobaric ensemble (NPT) at the same temperature. Uniaxial tensile strain applies then by a deformation control method. The strain applies every 200 time steps on the {\it x}-axis homogeneously with a strain rate of 0.0005 ps$^{-1}$. The strain rate is very small so we disregard viscous response, i.e. the system is given ample time to respond to the applied deformations. The Poisson's effect is accounted for by allowing the {\it y}-axis to relax during the slow elongation process. All of the MD simulations are performed using the LAMMPS (\cite{Plimpton1995}) software package.

As an outcome of this loading process, lenghts denoted by no. 1, 2, 3, 4 within the unit cell change. This change will not be same for all of them, due to their different position with respect to the loading direction. Pair of distances denoted by no. 1 tend to shorten since they are perpendicular to the loading direction; they are also affected equally due to loading, this is why we group them together. Pair of distances denoted by no. 2 tend to extent since they are inclined with respect to the extension direction. Certainly, due to symmetry and homogeneity of the applied extension, lengths denoted by no. 2 experience the same change. Pair of distances denoted by no. 4 tend to elongate by the same amount between them. For the distance denoted by no. 3 one expects elongation as well, since it is parallel to the direction of loading. 

Now, the idea is to correlate the continuum material parameters $c_1, c_2, c_5, c_9$ with distances denoted by no. 1, 2, 3, and 4 in the unit cell. This can be done using the following procedure: at zero strain level, we plot the radial distribution diagram (see Figure 2). 
\begin{figure}[!htb]
\centering
\includegraphics{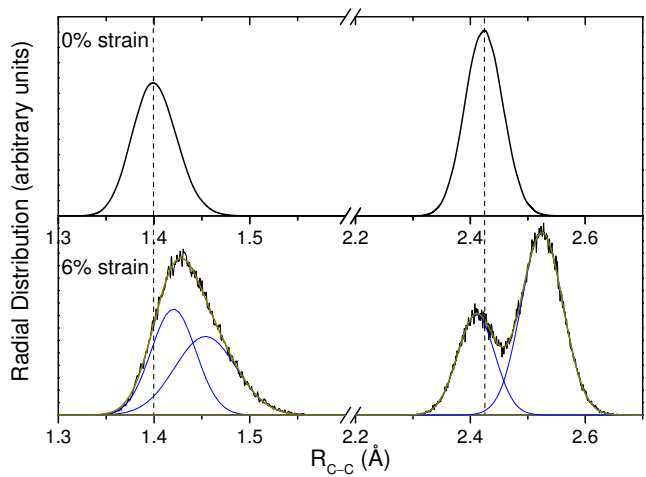}
\caption{ The radial distribution diagram at ease (above) and at strain level 6$\%$ (below). The horizontal axis measures carbon--carbon distances in Angstrom. }
\label{fig:digraph}
\end{figure}
On the horizontal axis of this diagram we have carbon--carbon distances measured in Angstrom. The radial distribution (or pair correlation) diagram describes the probability of encountering a carbon atom at any given distance from another carbon atom. To produce the radial distribution diagrams we extract atomic configuration at regular time step intervals. These are the same configuration that we later use for the calculation of the Poisson's ratio and Young modulus from the molecular modeling itself. For each atom we identify the first and second neighbors and calculate the corresponding distances. Thermal fluctuations alter these distances in a canonical (random) manner around a central value. Each of the distances is accounted for uniquely. In Figure 2 the formation of distinct density bands can be seen, one near the radial distance of 0.145 nm and another around 0.25 nm. These bands correspond to first and second nearest neighbors. At ease these band are Gaussian peaks, one centered at 0.242 nm and the second at 0.140 nm. These correspond to lengths of distances denoted by no. 1 and 2 and no. 3 and 4 at Figure 1, respectively. We remind here that at ease, pairs no. 1 and 2 have equal length of 0.242 nm. Also, pairs no. 3 and 4 have an equal length of 0.140 nm at ease. 

As loading applies these peaks split into two new peaks each (see Figure 2, bottom). The bottom plot of Figure 2 corresponds to the radial distribution at strain level 6$\%$. The splitting of each peak into two new ones is apparent. Essentially, these peaks describe the behavior of lengths no. 1--4 upon loading. Inspecting Figure 2 we see that at $6\%$ the peak centered at 0.242 splits into two new peaks: one at 0.240 nm and a second at 0.252 nm. These most probable values can be found by fitting two Gaussian functions to the band. As noted earlier, fitting with Gaussian functions is most appropriate due to the stochastic nature of the bond length variation which is a direct result of thermal movement. Each of the Gaussian functions correspond to specific lengths no. 1 and no. 2, and the centers (positions) of which provide the length values. Similarly, the peak centered at 0.140 nm at ease, upon strain splits into two peaks, one at 0.142 nm and one at 0.146 nm. Overall, this provides the evolution of lengths no. 3--4 upon application of strain.

Now, using these peaks we want to define the material parameters $c_1, c_2, c_5, c_9$. We repeat that the radial distribution diagram render the most probable lengths for no. 1--4 of Figure 1, after loading applies. We first define strain measures from the evaluated peaks as  
\begin{equation}
\textrm{strain \  measure \ for \ each \ peak}=\frac{\textrm{ ( final \ - \ initial ) \ value \ of \ the \ peak}}{\textrm{final \ value \ of \ the \ peak}}.
\end{equation}
Namely, for each peak value we subtract and divide by its initial value (namely the peak value for strain level zero). We then plot the applied stress (in absolute value) as a function of the above defined strain measures and obtain Figures 3--6. 
\begin{figure}[!htb]
\centering
\includegraphics{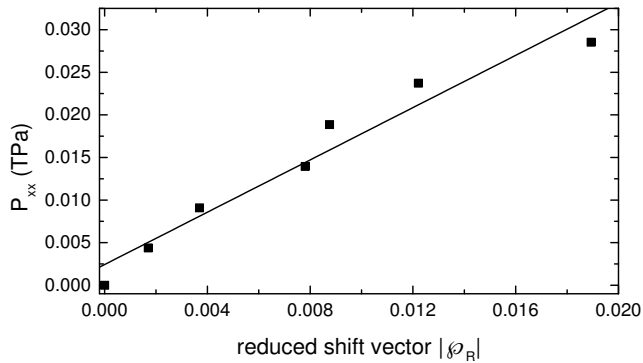}
\caption{Applied stress vs the strain defined by the left peak which initially was at 0.140 nm. The slope of the fitted line is 1.534. }
\label{fig:digraph}
\end{figure}
\begin{figure}[!htb]
\centering
\includegraphics{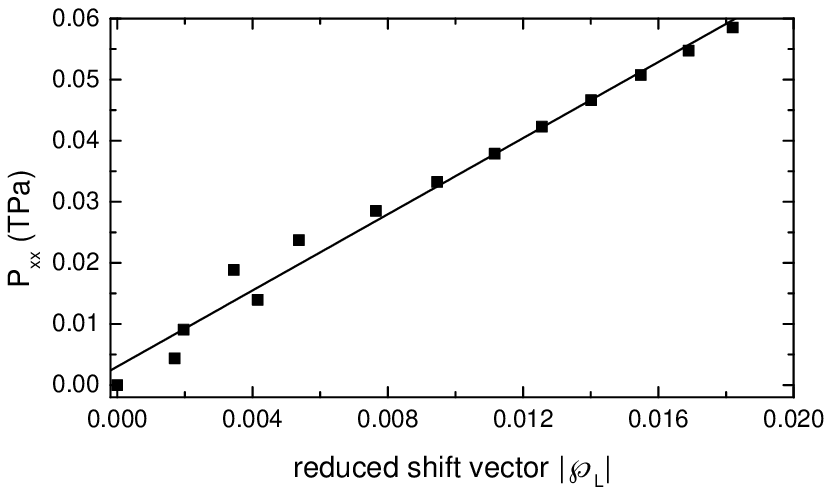}
\caption{Aplied stress vs the strain defined by the right peak which initially was at 0.140 nm. The slope of the fitted line is 3.117. }
\label{fig:digraph}
\end{figure}
\begin{figure}[!htb]
\centering
\includegraphics{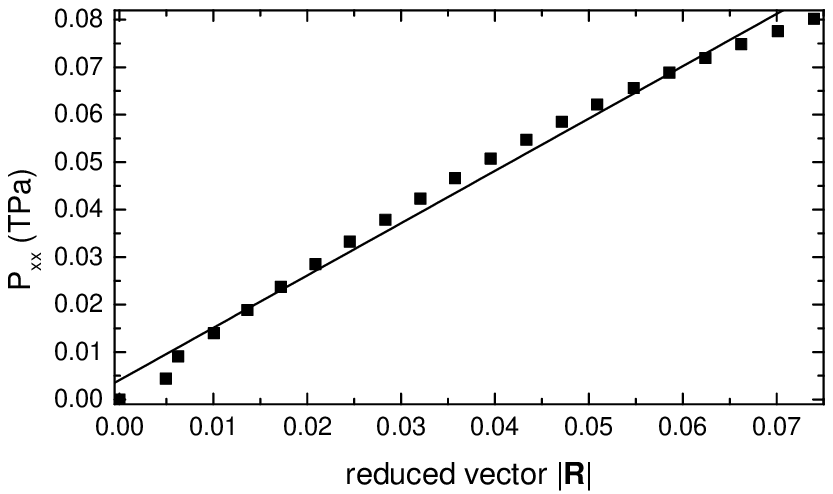}
\caption{Applied stress vs strain defined by the right peak which initially was at 0.240 nm. The slope of the fitted line is 1.102.  }
\label{fig:digraph}
\end{figure}
\begin{figure}[!htb]
\centering
\includegraphics{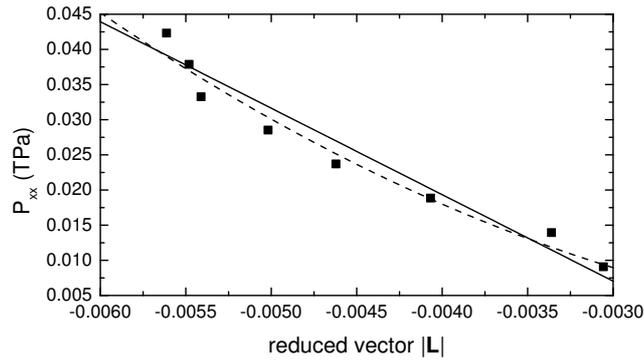}
\caption{Applied stress vs strain defined by the left peak which initially was at 0.240 nm. The dependence exhibits a parabolic form, nevertheless, not taking into account a small region near zero strain, the behavior 
is near linear with a slope of -12.7.  }
\label{fig:digraph}
\end{figure}
So, Figures 3--6 render how carbon--carbon distances denoted by no. 1--4 in Figure 1 change with applied loading when one can do calculations at the molecular level. Figure 3 relate the applied stress with the increment of the left peak when initially the peak is at 0.140 nm; we call this left peak strain measure as the reduced shift vector ${\mathcal P}_R$. Figure 4 plot the increment in the right peak when initially the peak is at 0.140 nm with applied stress; we call the strain measure calculated from the right peak as reduced shift vector ${\mathcal P}_L$. Figures 5 and 6 are analogous for the peak which initially was at 0.240 nm; the corresponding strain measures are denoted as reduced vectors $|{\bf R}|$ and $|{\bf L}|$, respectively. 

One remark is in order here regarding the definition of the applied stress $P_{xx}$. The whole process is deformation controlled. Nevertheless, one may convert the applied strain to stress using the manual convertion of the LAMMPS programs, which is based on the use of the virial theorem. This renders a three dimensional definition of applied stress. To projected this quantity to graphene's sheet one has to divide this three dimensional stess by graphene's thickness, taken to be approximately 0.335 nm. 

We now define the material parameters $c_1, c_2, c_5, c_9$ as the slope of diagrams 3-6. The outcome regarding the sorresponding slopes render values $3.117, 1.534, 1.102, -12.07$ in TPa. Inspecting these diagrams one can see the different behaviour of Figure 6 from the rest figures. The slope of diagram 6 is negative, while for the rest figures render a positive slope. This is physically reasonable since we expect length no. 1 in Figure 2 to shorten. Also, it has a parabolic character, nevertheless, not taking into account a small region near zero strain, the behaviour is near linear, as is seen in Figure 6.   

Now, what it remains to be done is to juxtapose lengths no. 1--4 to material parameters $c_1, c_2, c_5, c_9$. At a first look it seems reasonable to associate $c_1, c_2$ with 1 and 2 and $c_5, c_9$ with 3 and 4, namely to have the fourtuple $(c_1, c_2, c_5, c_9)=(-12.07, 1,102, 3.117, 1.534)$ in TPa. This is mainly due to the fact that diagrams of Figures 3 and 4 plot the peaks produced by the initially 0.140 nm peak, namely terms related with the shift vector at the unit cell. Similarly, Figures 5 and 6 plot the peaks produced by the initially 0.242 nm peak, namely terms related with the lattice vectors at the unit cell. But this expectation is not necessarily true due to the following reason. At the discrete level one views the unit cell and distinguishes between distances 1, 2, 3 and 4. When one passes to the continuum, the continuum analogue of the unit cell "patches" at only one continuum point. Thus, information regarding the atomic level "patch" to only one material point and it's four material parameters. This is at the very root of the multiscale method and certainly some information is lost.  

So, when one "scales up" (see Figure 7)
\begin{figure}[!htb]
\centering
\includegraphics{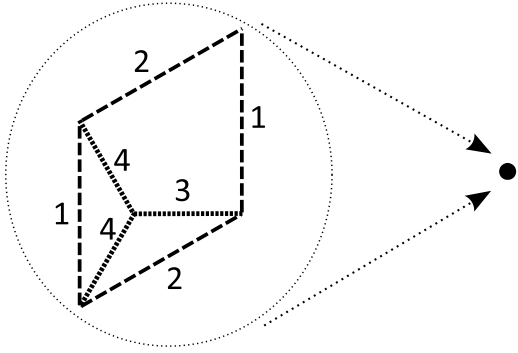}
\caption{On the left we see the disrete view within the unit cell of graphene. On the right we have the continuum analog, which is a continuum point with material parameters $c_1, c_2, c_5, c_9$. }
\label{fig:digraph}
\end{figure}
to the continuum level loses the ability to distinguish which distance in the unit cell corresponds to the material parameters $c_1, c_2, c_5, c_9$ pertain. Further discussion on the subject is presented in the following section where validation comparing with standard measurement is presented. There we see that the expectation of corresponding lenghts no. 1, 2, 3, 4 of Figure 1 to $c_1, c_2, c_5, c_9$ is not borne out, when we take as a minimum requirement to have values for $(E, \nu)$ comparable to the one's from literature. 

To sum up, we define material parameters $(c_1, c_2, c_5, c_9)$ as the slope in the diagrams of applied stress versus the four newly defined strain measures. These new strain measures pertain to length changes denoted as numbers 1--4 in Figure 1. The values are then $-12.07, 1.102, 3.117, 1.534$ in TPa.

\section{Correlation with well-accepted measurements}

As a method of validation/calibration of our theory, we compare our calculated values with some of the most well-accepted measurements from the vast literature on the topic (see the Introduction section for relevant citations). This can also help as a guideline for making the correct association between material parameters $c_1, c_2, c_5, c_9$ and the evaluated slopes. To obtain the relation between $(E, \nu)$ and $(c_1, c_2, c_5, c_9)$ we start by solving the equations ruling the shift vector: these are eqs. (11). From eq. (8), setting the right hand side equal to zero, we obtain
\begin{equation}
2 c_5 p_1=\frac{4 c_5^2 e_{12}}{c_9}.
\end{equation}
Eq. (9), with a zero on the right hand, solves to give
\begin{equation}
c_5 p_2=\frac{c_5^2 (e_{11}-e_{22})}{c_9}.
\end{equation}
Substituting these expressions to eqs. (3--5) we obtain in matrix form
\begin{equation}
\begin{pmatrix}
  \sigma_{11}  \\
  \sigma_{22}  \\
  \sigma_{12}  \\
\end{pmatrix}=
\begin{pmatrix}
  c_1-\frac{c_5^2}{c_9} & \ \ c_2+\frac{c_5^2}{c_9} & \ \ 0 \\
  c_2+\frac{c_5^2}{c_9} & \ \ c_1-\frac{c_5^2}{c_9} & \ \ 0 \\
  0 & \ \ 0 & \ \ \frac{c_9(c_1-c_2)-8c_5^2}{2 c_9} \\
\end{pmatrix}
\begin{pmatrix}
  e_{11}  \\
  e_{22}  \\
  e_{12}  \\
\end{pmatrix}.
\end{equation}

Inversion of the above relations render in matrix form
\begin{equation}
\begin{pmatrix}
  e_{11}  \\
  e_{22}  \\
  e_{12}  \\
\end{pmatrix}=
\begin{pmatrix}
  \frac{\alpha}{\alpha^2-\beta^2} & \ \ -\frac{\beta}{\alpha^2-\beta^2} & \ \ 0 \\
  -\frac{\beta}{\alpha^2-\beta^2} & \ \ \frac{\alpha}{\alpha^2-\beta^2} & \ \ 0 \\
  0 & \ \ 0 & \ \ \frac{1}{c} \\
\end{pmatrix}
\begin{pmatrix}
  \sigma_{11}  \\
  \sigma_{22}  \\
  \sigma_{12}  \\
\end{pmatrix},
\end{equation}
where $\alpha=c_1-\frac{c_5^2}{c_9}$, $\beta=c_2+\frac{c_5^2}{c_9}$ and $c= \frac{c_9(c_1-c_2)-8c_5^2}{2 c_9}$. Setting all stress tensor components equal to zero except $\sigma_{11}$ we obtain that 
\begin{equation}
e_{11}=\frac{\alpha}{\alpha^2-\beta^2} \sigma_{11}.
\end{equation}

Since for the linear case Young's modulus is defined as $e_{11}=\frac{1}{E} \sigma_{11}$, we obtain for our case
\begin{equation}
E=\frac{\alpha^2-\beta^2}{\alpha}=\frac{[c_1-\frac{c_5^2}{c_9}]^2-[c_2+\frac{c_5^2}{c_9}]^2}{c_1-\frac{c_5^2}{c_9}}.
\end{equation}
Poisson ration is defined for our framework as
\begin{equation}
\nu=-\frac{e_{22}}{e_{11}}=-\frac{-\frac{\beta}{\alpha^2-\beta^2}}{\frac{\alpha}{\alpha^2-\beta^2}}=\frac{\beta}{\alpha}=\frac{c_2+\frac{c_5^2}{c_9}}{c_1-\frac{c_5^2}{c_9}}.
\end{equation}
The last two equations are the connection between our mechanical approach with the standard material parameters of the linear modeling of graphene, at small strains. It is obvious that material parameters $c_1, c_2, c_3, c_4$ work synergetically to produce the Young modulus and the Poisson ratio. One cannot attribute changes of length to the direction of loading to only one of the material parameters $c_1, c_2, c_5, c_9$. The same holds true for changes in length along the direction perpendicular to loading.  Also, it is obvious that one cannot invert eqs. (18, 19) to solve in a unique way for $c_1, c_2, c_5, c_9$ as functions of $(E, \nu)$. This is expected since the present framework has four material parameters while classical linear elasticity has only two.  

This should not be confused with the inversion procedure for the passage from eq. (15) to eq. (16). There, we solve the equation ruling the shift vector, eq. (11), to obtain the shift vector as a function of the strain components (see eqs. (13, 14)). Due to the fact that we use a linear theory, the relation between the shift vector components and the strain components is linear. Having the shift vector components as linear functions of the strain components we substitute them to the constitutive law (eqs. (3-5)). This way we ontain eq. (15), which is a non-standard constitutive law (since it contains $c_1, c_2, c_5, c_9$) relating stresses to strains. Even though it is non-standard it is linear. Thus, it can be inverted giving eq. (16).  

To find the appropriate values of $c_1, c_2, c_5, c_9$ we choose from the pool of values $-12.07, 1.102,$ $3.117, 1.534$ in TPa and substitute them to eqs. (18, 19). The choice $(c_1, c_2, c_5, c_9)=(1.102, 1.534, 3.117, -12.07)$ is the optimum, in the sense of having values of $(E, \nu)$, calculated from eqs. (18, 19), as close to the one reported in literature as possible. It is clear that the outcome values of $(E, \nu)=(1.37 \text{ TPa}, 0.41)$ overestimate both measures, but nevertheless, remains within the range of reasonable values for these measures.  

On the other hand, for the calculation of the Poisson's ratio and Young modulus from the molecular dynamics simulations solely (namely, without the need of introducing $c_1, c_2, c_5, c_9$), atomic configurations are extracted from the simulation trajectory at regular time step intervals that correspond to regular increases in applied strain. The calculated values are $(E, \nu)$=(0.95 TPa, 0.20). Care is taken so that the structure has sufficient time to equilibriate following the latest deformation (strain increase). The Poisson's ratio is calculated by measuring the changes of both the lateral and transverse dimensions of the computational cell. Using the same information the corresponding strain levels are recorded as well. For the Young modulus, in addition to the previous, the corresponding applied pressures $P_{xx}$ at the graphene edges are also needed, of course suitably scaled to account for the difference between height of the computational cell and thickness of graphene. The Young modulus is then calculated as the slope of the pressure--strain diagram at small strains ($ < 6\%$).

Values for E and $\nu$ from the reported literature (see the first four paragraphs of this Section) range as E=0.67-1.91 TPa and $\nu$=0.14-0.45 depending on the methodology used. The central tendency of these values for E is the value 1 $\pm$ 0.15 TPa. Compared to this value our outcome of $(E, \nu)$=(1.37 TPa, 0.41) overestimates these quantities but still remain within the range of acceptable values. From the literature cited, the continuum methods (i.e. the finite element approaches of \cite{Arroyo-Belytschko2004,Reddyetal2005,Scarpaetal2009}) tend to have greater discrepancy from the value 1 $\pm$ 0.15 TPa. Thus, our theory being ultimately a continuous theory is expected to follow this trend. 

On the other hand, our pure molecular mechanics modeling render values $(E, \nu)$=(0.95 TPa, 0.20). These values are obtained using the definition of the AIREBO manual. On the other hand, the values $(E, \nu)$=(1.37 TPa, 0.41) are based on a different definition of $(E, \nu)$: they are based on a genuinely continuous definition which is non-standard since it uses $c_1, c_2, c_5, c_9$. Certainly, the two definitions (the discrete and the continous one) measure the same quantities in a different way. So, the discrepancy in their reported values is based on the different definition but still remains in the range of an admissible difference.  

So, all in all, from molecular calculations we determine/define the following values for the material parameters $(c_1, c_2, c_5, c_9)=(1.102, 1.534, 3.117, -12.07)$ in TPa. These are the material parameters needed when graphene is modeled as a hexagonal 2-lattice at the continuum level. They appear to the non-standard constitutive law (eqs. (3-5)) and characterize the stress-strain response in this case. Ultimately, they lead to values  $(E, \nu)=(1.37 \text{ TPa}, 0.41)$ through eqs. (18, 19); these are slightly overestimated values which nevertheless, remain in the range of reasonably accepted values. 

\section{Conclusion and future directions}

The present work involves a molecular study with the purpose of measuring in-plane material moduli for graphene at the continuum level. The theoretical framework adopted is restricted to material and geometrical linearities. Graphene is modeled as a hexagonal 2-lattice, so for the linear regime there are four material parameters for in-plane motions. We evaluate these material parameters using molecular mechanics and the AIREBO potential. The material parameters are defined as the slopes of stress--strain diagrams of suitably defined strain measures from graphene's unit cell at the discrete level. The final values evaluated are  $(c_1, c_2, c_5, c_9)=(1.102, 1.534, 3.117, -12.07)$ in TPa and correspond to Young modulus and Poisson ratio $(E, \nu)=(1.37 \text{ TPa}, 0.41)$. 

The future direction of the authors regarding this problem is further exploration of the nonlinear counterpart of the present theoretical framework, with the purpose of capturing the effects of large strains on free-standing graphene monolayers. In that case even though the approach will be similar with the one presented, several difficulties arise. Firstly, the complexity of the model for the genuinely nonlinear is much greater, since that model involves nine material parameters for in-plane motions only. Secondly, the strains and stresses will be much higher than the small strains used here; this requires a much more demanding set of molecular mechanics calculations for loadings up to 24$\%$. Thirdly, simple shear as well as pressure computational experiments should be used in addittion to the axial extension program utilized here. As in the linear case studied here, for the nonlinear case as well we distinguish between the shift vector components and the lattice vector components and define the nonlinear material parameters as changes of these components against suitable loading.  

All in all, we believe that our strict theoretical modeling can capture all the interesting phenomena that occur during the loading of monolayer-thick graphene sheets. Properly designed molecular mechanics simulations, with the AIREBO potential, can provide the material moduli introduced by the theoretical modeling, and enables verification of the results with experimental values. For the case of small strains presented here, we are able to reproduce with reasonable good agreement many of the experimental/computational results, and generalization of the method for large strains will follow in future works. \\

{\it{Acknowledgments}}\\

We thank G.I. Sfyris (Athens, Greece) for reading the manuscript throughout and for numerous most valuable addittions in the draft. This research has been co-financed by the European Union (European Social Fund -- ESF) and Greek national funds through the Operational Program "Education and Lifelong Learning" of the National Strategic Reference Framework (NSRF) Research Funding Program: ERC-10 "Deformation, Yield and Failure of Graphene and Graphene-
based Nanocomposites". The financial support of the European Research Council through the projects ERC AdG 2013 (‘‘Tailor Graphene’’) is greatfully acknowledged. The authors would like to acknowledge the financial support of Graphene FET Flagship (Graphene-Based Revolutions in ICT and Beyond, Grant No. 604391N.M.P. is supported by the European Research Council (ERC StG Ideas 2011 BIHSNAM no. 279985 on ‘Bio-inspired hierarchical supernanomaterials’, ERC PoC 2013-1 REPLICA2 no. 619448 on ‘Large-area replication of biological anti-adhesive nanosurfaces’, ERC PoC 2013-2 KNOTOUGH no. 632277 on ‘Super-tough knotted fibres’), by the European Commission under the Graphene Flagship (WP10 ‘Nanocomposites’, no. 604391) and by the Provincia Autonoma di Trento (‘Graphene nanocomposites’, no. S116/2012-242637 and reg. delib. no. 2266).



\vspace{0.1cm}

D. Sfyris\\
Foundation for Research and Technology \\
Institute of Chemical Engineering Sciences, Patras, Greece \\
dsfyris@iceht.forth.gr \\
dsfyris@sfyris.net \\

\vspace{0.1cm}

E.N. Koukaras\\
Foundation for Research and Technology \\
Institute of Chemical Engineering Sciences, Patras, Greece\\
Department of Physics, University of Patras, Greece \\

\vspace{0.1cm}

N. Pugno \\
Laboratory of Bio-Inspired and Graphene Nanomechanics, Department of Civil,
Environmental and Mechanical Engineering, Universita` di Trento, via Mesiano, 77,
38123 Trento, Italy \\
Center for Materials and Microsystems, Fondazione Bruno Kessler, Via Sommarive 18,
38123 Povo (Trento), Italy \\
School of Engineering and Materials Science, Queen Mary University of London, Mile
End Road, London E1 4NS, UK \\

\vspace{0.1cm}

C. Galiotis \\
Foundation for Research and Technology \\
Institute of Chemical Engineering Sciences, Patras, Greece\\
Department of Chemical Engineering, University of Patras, Greece

\end{document}